\documentclass[pra,preprint]{revtex4-1}
\usepackage{graphicx}
\usepackage{epstopdf}
\newcommand{\dk}{{{\mit \Delta} k}}

\newcommand{\dt}{{{\mit \Delta} t}}
\newcommand{\e}{{\rm e}}
\newcommand{\al}{{\alpha}}

\newcommand{\ld}{{\lambda}}

\newcommand{\pa}{{\partial}}

\newcommand{\bea}{\begin{eqnarray}}
\newcommand{\eea}{\end{eqnarray}}
\newcommand{\be}{\begin{equation}}
\newcommand{\ee}{\end{equation}}
\newcommand{\ba}{\begin{eqnarray}}
\newcommand{\ea}{\end{eqnarray}}

\newcommand{\nn}{\nonumber}
\newcommand{\la}{\label}
\newcommand{\w}{\Omega}

\begin{document}
\title{Anatomy of the Akhmediev breather: cascading instability, first formation time and Fermi-Pasta-Ulam recurrence}

\author{Siu A. Chin$^\dagger$, Omar A. Ashour$^\ddagger$, Milivoj R. Beli\'c$^\ddagger$ }
\affiliation{$^\dagger$Department of Physics and Astronomy,
Texas A\&M University, College Station, TX 77843, USA}
\affiliation{$^\ddagger$Science Program, Texas A\&M University at Qatar,
P.O. Box 23874 Doha, Qatar}

\begin{abstract}
By invoking Bogoliubov's spectrum, 
we show that for the nonlinear Schr\"odinger equation, the modulation instability of
its $n=1$ Fourier mode on a finite background automatically triggers a further cascading instability,
forcing all the higher modes to grow exponentially in locked-step with the $n=1$ mode.
This fundamental insight, the enslavement of all higher modes to the $n=1$ mode,
explains the formation of a triangular-shaped spectrum which generates the Akhmediev breather,
predicts its formation time analytically from the initial modulation amplitude,
and shows that the Fermi-Pasta-Ulam (FPU) recurrence is just a matter of energy conservation with
a period twice the breather's formation time. 
For higher order MI with more than one initial unstable modes, while most evolutions
are expected to be chaotic, we show that it is possible to have isolated cases of
``super-recurrence'', where the FPU period is much longer than that of a single
unstable mode.

\end{abstract}

\pacs{xxx}

\maketitle
\section{Introduction}

The study of modulation instability (MI) in solutions of the nonlinear Schr\"odinger equation on a constant 
background has became a cornerstone of modern nonlinear physics, underlying many of the advances in understanding deep water wave
propagation \cite{ben67,lak77,yue78}, plasma physics \cite{tan68}, light transmission in optical fibers \cite{tai86}, 
and the formation of optical rogue waves \cite{solli07,dudley14}.
While directly solving the nonlinear Schr\"odinger numerically is a relatively simple matter \cite{yue78},
exact solutions to the nonlinear Schr\"odinger equation on a finite background, known as the 
Akhmediev \cite{akh86,akh87} and Kuznetsov-Ma \cite{kuz77,ma79} breathers (ABs, KMBs),
have provided much insight into the subsequent evolution of MI in these solutions.
In this work, we seek to provide a more detailed understanding of the
Akhmediev breather's formation, its fundamental structure and its
remarkable mode of evolution.

Akhmediev breathers \cite{akh86,akh87} are exact solutions to the cubic nonlinear Schr\"odinger
\be
i\frac{\partial\psi}{\partial t}
+\frac12 \frac{\partial^2\psi}{\partial x^2}+|\psi|^2\psi=0
\la{sch}
\ee
on a finite background, $|\psi(t \rightarrow\pm\infty)|\rightarrow 1$.
In this work, we show that: 1) The modulation instability of
the $n=1$ Fourier mode, automatically triggers a further {\it cascading instability},
forcing all the higher modes to grow exponentially in locked-step with the $n=1$ mode.
This results in a {\it triangular spectrum} \cite{akh11}, which is {\it the} signature of the Akhmediev breather.
The remarkable simplicity of the Akhmediev breather is that, once formed, this triangular spectrum
basically evolves intact, throughout its subsequent evolution, oblivious to any
nonlinear interactions. 2) By knowing the analytical form of the Fourier amplitudes
from AB, one can predict the time of the breather's first formation.
This formation time corresponds to the maximum compression distance \cite{erk11a} in
an optical fibre, and is an important design parameter for breather productions. 
Our formation time is an improvement over that derived in Ref.\cite{erk11a};
it is accurate even for purely real modulation amplitudes.
3) Since all higher Fourier modes are ``enslaved" \cite{inf81} to the $n=1$ mode, there is
no freedom for the equal partition of energy and no Fermi-Pasta-Ulam \cite{fer65,ford92} (FPU) paradox.
The FPU recurrence is then just a consequence of bound state energy conservation with a period
twice the breather's formation time. 4) In cases of higher-order modulated instabilities, where there
are multiple initial unstable modes, super FPU recurrences are possible, but Fourier amplitudes
beyond the first AB-like peak are no longer predicted by the Akhmediev breather.

\section{Anatomy of the Akhmediev breather}

      The Akhmediev breather \cite{akh86,akh87} 
\be
\psi(t,x)=\frac{(1-4a)\cosh(\ld t)+\sqrt{2a}\cos(\Omega x)+i\ld\sinh(\ld t)}{\sqrt{2a}\cos(\Omega x)-\cosh(\ld t)},
\la{ab}
\ee
is an exact solution to (\ref{sch}) parametrized by a single real positive parameter $a$, which fixes the wave number $\w$ and the growth factor $\ld$. In order to make clear the most fundamental aspect of the solution, it is best to regard $a$
as parametrizing the solution's periodic length $L$:
\be
L=\frac{\pi}{\sqrt{1-2a}}.
\la{len}
\ee
For an AB, $a$ ranges between 0 and 1/2; at $a=1/2$ the Peregrine soliton \cite{per83} forms, with an infinite periodic length.
Given $L$, the spacing in $k$-space is
 $
 \dk={2\pi}/{L},
 $
 so that the allowed $k$ vectors are just
 \be
 k_n=n\dk\qquad{\rm for}\quad n=0,\pm1,\pm2, \cdots .
 \ee
 The wave number $\Omega$ of the Akhmediev breather (\ref{ab}) then corresponds to the
 fundamental, $n=1$ mode
 \be
 \Omega=\dk=2\pi/L=2\sqrt{1-2a},
 \ee
and the growth factor
\be
\ld=\sqrt{8a(1-2a)},
\ee
is due to the instability of this mode, as determined by the Bogoliubov spectrum \cite{bog47}.

 While the general Benjamin-Feir \cite{ben67} instability is known since 1967, the modulation instability of
the cubic nonlinear Schr\"odigner equation is known from Bogoliubov's work on the uniform Bose
gas \cite{bog47,fet71} since 1947. This is because a uniform Bose gas can be described by the Gross-Pitaevskii
equation \cite{fet71,pet02}, which is just the cubic nonlinear Schr\"odinger equation with a uniform background.

Bogoliubov's spectrum \cite{bog47,fet71} for the elementary
excitations of a uniform Bose gas is given by
\be
\varepsilon_k=\sqrt{E_k(E_k+2U)}
\la{bs}
\ee
where $E_k=k^2/2$ is the free-particle energy and $U=g|\psi|^2$.
In the repulsive (defocusing) case of $g=+1$, all elementary excitations are
stable. In the attractive (focusing) case of $g=-1$, all $k$-modes with
plane-wave $e^{ikx}$ are unstable if $E_k+2U<0$. In the latter case,
for a constant background $|\psi|^2=1$, the
$n=\pm 1$ modes are unstable with imaginary frequencies
 \ba
 \varepsilon_{\pm 1}&=&\pm i \sqrt{ \frac{\dk^2}{2} ( 2-\frac{\dk^2}{2} )}=\pm i \sqrt{8a(1-2a)}=\pm i \ld,
 \ea
 and the modulus of the amplitudes $A_{\pm 1}$ grows in time as
 \be
|A_{\pm 1}|\propto \e^{\mp i\varepsilon_{\pm 1} t}=\e^{\ld t}.
\la{aone}
 \ee
Thus, the instability of the $n=\pm1$ modes determines the growth factor $\ld$.
More generally, since a $k$-mode is unstable for $E_k-2<0$, this means that
all modes with $k_n<2$, or $n\w<2$, are unstable. For $1<\w<2$ only
one mode is unstable. If $\w<1$, then there will be more than one unstable
mode with more than one growth factor. The case
of $\w=1$ corresponds to $a=3/8=0.375$, hence there will be multiple unstable
modes initially if $a>0.375$.

\begin{figure}[hbt]
\includegraphics[width=0.95\linewidth]{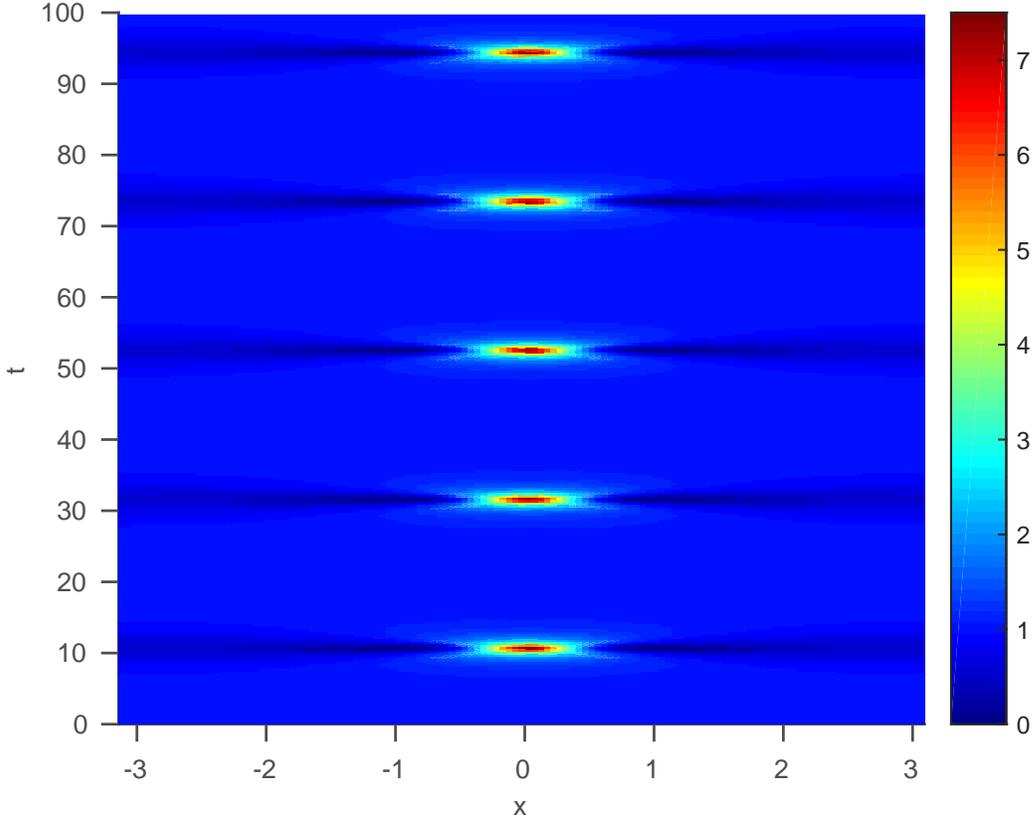}
\caption[]{\label{density} (color online)
  Density $|\psi(t,x)|^2$ plot of the numerical solution of the cubic nonlinear Schr\"odinger equation (\ref{sch})
  using a second-order splitting method with $\dt=0.0001$ for $a=3/8$ and initial profile (\ref{init}).
}
\end{figure}

 When starting with a constant background of $\psi=1$, with $A_0=1$ and $A_{n\ne 0}=0$,
 any minute perturbation which triggers the instability of the $n=1$ mode will cause it to grow exponentially,
 according to (\ref{aone}). This is the standard Benjamin-Feir \cite{ben67} scenario.
 What has not been explicitly stated prior to this work is that, for the nonlinear Schr\"odinger equation,
 this growth of the $n=1$ mode will {\it automatically} trigger a {\it cascading instability} of all the higher modes, 
 causing all to grow exponentially, locked to the fundamental mode. 
 This is because the Bogoliubov spectrum is obtained by linearizing
 \be
 |\psi|^2=|A_0+A_1\e^{i\dk x}+A_{-1}\e^{-i\dk x}|^2
 \ee
 to leading orders in $A_{\pm 1}$. The final equation is
 of the form
 \be
 i\pa_t A_{\pm 1}\propto A_{\pm 1},
 \ee
 which results in an exponential growth factor. Doing the same expansion for
 $A_{\pm 2}$ in
 \be
 |\psi|^2=|A_0+A_1\e^{i\dk x}+A_{-1}\e^{-i\dk x}+A_2\e^{i2\dk x}+A_{-2}\e^{-i2\dk x}|^2
 \ee
 now yields leading order contributions of the form
 \be
  i\pa_t A_{\pm 2}\propto A_{\pm 1}^2A_0^*+2A_0A_{\pm 1}A_{\mp 1}^*,
 \ee
 which are simply proportional to a product of two $A_{\pm 1}$. This is because the amplitudes $A_{\pm 2}$ are
 just starting to grow from zero, and are much smaller than the constant $A_0\approx 1$ and the already growing $A_{\pm 1}$. Thus $A_{\pm 2}$ are given by a simple {\it time integration},
 \be
  A_{\pm 2}\propto \int (A_{\pm 1}^2A_0^*+2A_0A_{\pm 1}A_{\mp 1}^*) dt
 \ee
 resulting in
  \be
  |A_{\pm 2}|\propto |A_{1}|^2\propto \e^{2 \ld t}.
 \ee
 Repeating similar argument for $A_{\pm n}$ ($n\ne 0$) then yields
 \be
  |A_{\pm n}|=C_n\e^{|n|\ld t}=C_n|A_{1}|^{|n|}.
  \la{ana}
 \ee
Therefore the growth of the entire spectrum is dictated by the growth of $|A_{1}|$.
If $C_n$ does not grow exponentially faster than $n$,
then at $t<0$, $\ln(|A_{\pm n}|)\propto -|n|$, which is triangular-shaped spectrum \cite{akh11} in $n$.

\begin{figure}[hbt]
\includegraphics[width=0.95\linewidth]{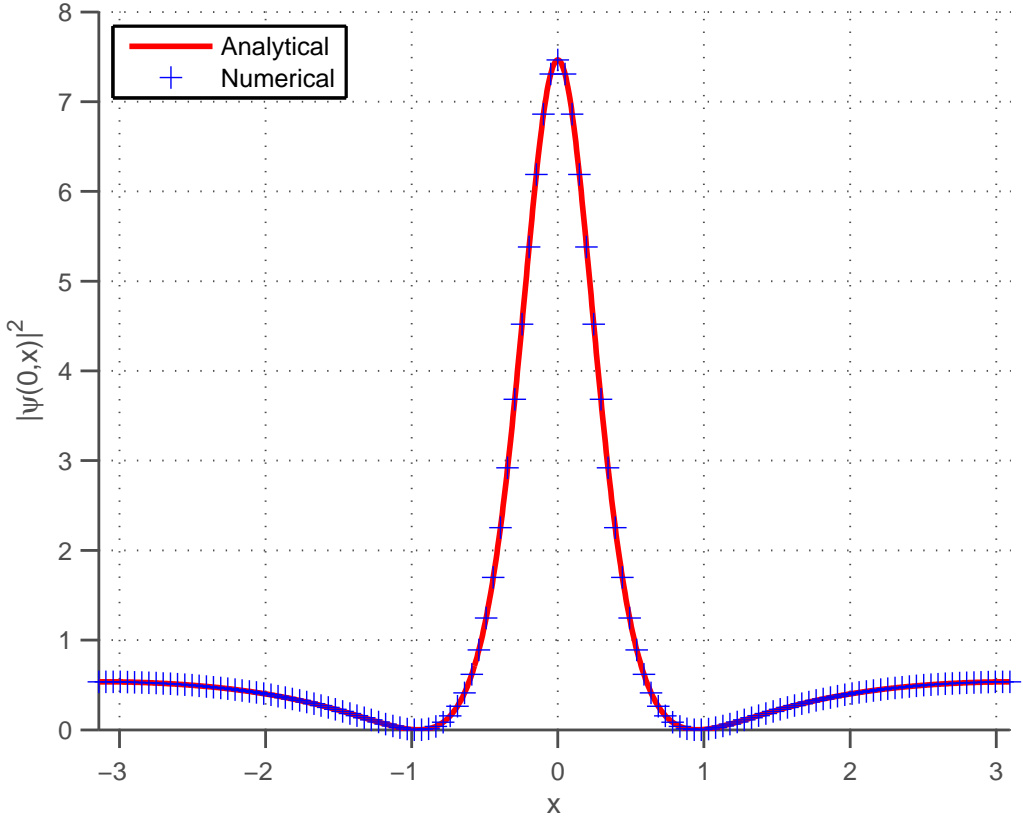}
\caption[]{\label{spatial} (color online)
The first breather's numerical spatial profile from the calculation of Fig. \ref{density}
as compared to the analytical form of $|\psi(0,x)|^2$ from Eq. (\ref{ab}).
}
\end{figure}

\begin{figure}
\includegraphics[width=0.95\linewidth]{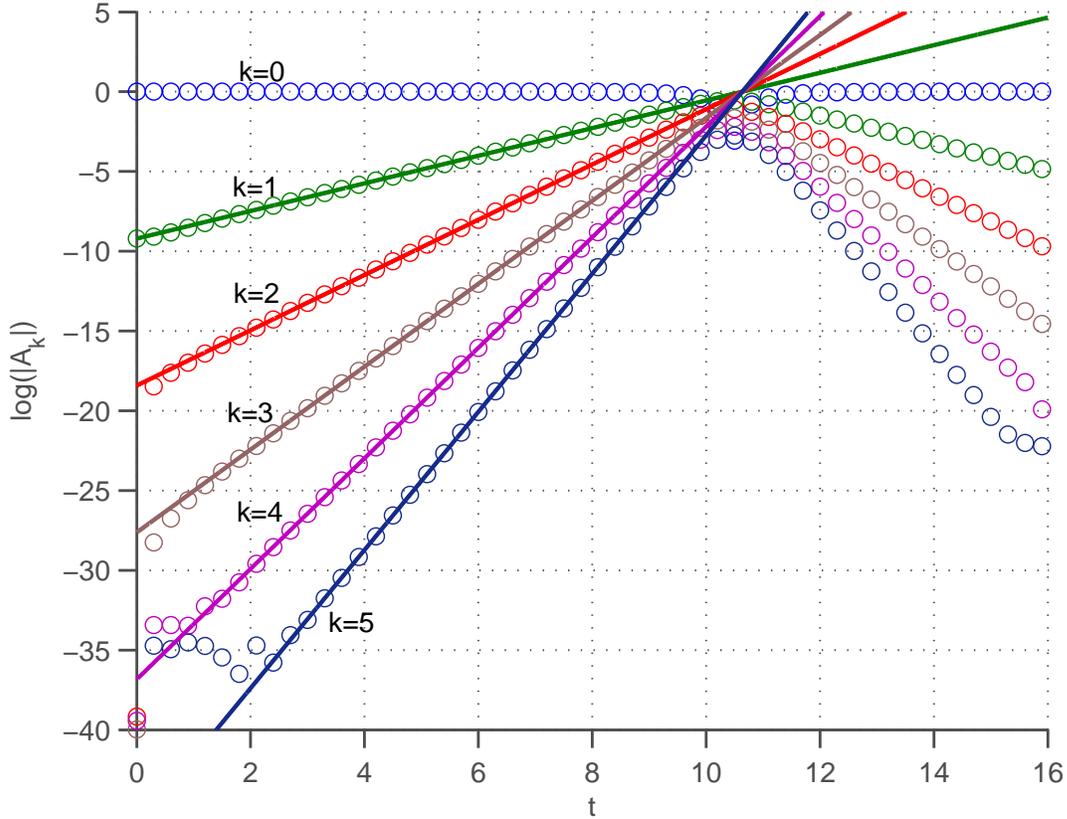}
\caption{ The growth of the $k=1-5$ Fourier amplitudes in time. Symbols are numerical results 
from the calculation of Fig. \ref{density}.
Solid lines are $k\ld (t-t_c)$ with $\ld=\sqrt{3}/2$ and $t_c=10.6352$.
\label{kmode}}
\end{figure}

 To check the validity of this cascading scenario, we solve the
 nonlinear Schr\"odinger equation numerically using a second-order splitting
 method for the case of  $a=3/8$, $L=2\pi$, $\dk=1$, $\w=1$, $\ld=\sqrt{3}/2$,
with an initial profile
\be
\psi(0,x)=A_0+A_1\e^{i\w x}+A_1\e^{-i\w x}=A_0+2A_1\cos(\w x),
\la{init}
\ee
where
$A_0=\sqrt{1-2A_1^2}\quad{\rm and}\quad A_1=10^{-4},
$
so that wave function is normalized as in Ref. \cite{akh86}:
\be
\frac1{L}\int_0^{L}|\psi(0,x)|^2 dx=A_0^2+2A_1^2=1.
\ee
The allowed $k$-modes are just integers $k=n=0,\pm1,\pm2,$ {\it etc.}.
In Fig. \ref{density}, the resulting density profile plot shows that the breather is first formed at $t\sim 10$
and then recurs later at intervals of $t\sim 20$. To verify that the structure formed
is precisely the breather of (\ref{ab}) with $a=3/8$, we compare in Fig. \ref{spatial},
the structure's spatial profile at the formation time with
$|\psi(0,x)|^2$ of Eq.(\ref{ab}). The agreement is exact.

In Fig. \ref{kmode}, the growth of $|A_{k}|$ for $k=1-5$
is compared to the cascading prediction (\ref{ana}) that $|A_{k}|=\e^{k\ld (t-t_c)}$, with
the prefactor $C_n$ absorbed into the shift of the time origin $t_c$.
All $k=1-5$ amplitudes can be well-fitted with a single cascading time of $t_c=10.6352$,
\be
\ln(|A_{k}|)=|k|\ld(t-t_c),
\la{lna}
\ee
where $t_c$ is the time needed for $A_{1}$ to grow
from an initial value of $10^{-4}$ to unity at the rate of $\ld=\sqrt{3}/2$,
\be
t_c=-\frac{\ln(A_1)}{\ld}=10.6352.
\la{tc}
\ee
The excellent linear fits to the amplitudes in Fig. \ref{kmode} are therefore parameter-free
predictions of the cascading instability.
It only fails to describe the growth of the amplitudes
near and after the amplitudes' peak, at which time the exact dynamics takes over. 
The breather's actual formation time of
\be
t_0=10.4691
\la{tf}
\ee
will be derived in a later section; the cascading time $t_c$ is just a first estimate of $t_0$. 
Because of the cascading instability, the growth of the amplitudes is best understood
and plotted in terms of $\ln(|A_{k}|)$ rather than $|A_{k}|$ or $|A_{k}|^2$.

Of course, this exponential growth in the amplitudes cannot continue indefinitely, since it must be
constrained by the unitary condition on the wave function:
 \be
|A_0|^2+ 2\sum_{n>1}|A_n|^2=1.
 \ee
 If one assumes that the cascading spectrum (\ref{ana}) persists (setting $C_n=1$) even when $|A_1|$ can no longer be considered as ``small'', then since $|A_0|$ can only be depleted to zero, the maximum that $|A_1|$ can grow to is given by
 \be
\sum_{n>1}(|A_1|^2)^n=\frac12\quad\Rightarrow\quad max(|A_1|)=\frac1{\sqrt{3}}
 \ee
 and $max( |A_n|)=(1\sqrt{3})^n$. As we will see in the later sections, after the amplitudes have
 reached their maxima, by energy conservation, they must decline back to their starting values,
 in a time-symmetric image of their rise.
 
\begin{figure}
\includegraphics[width=0.95\linewidth]{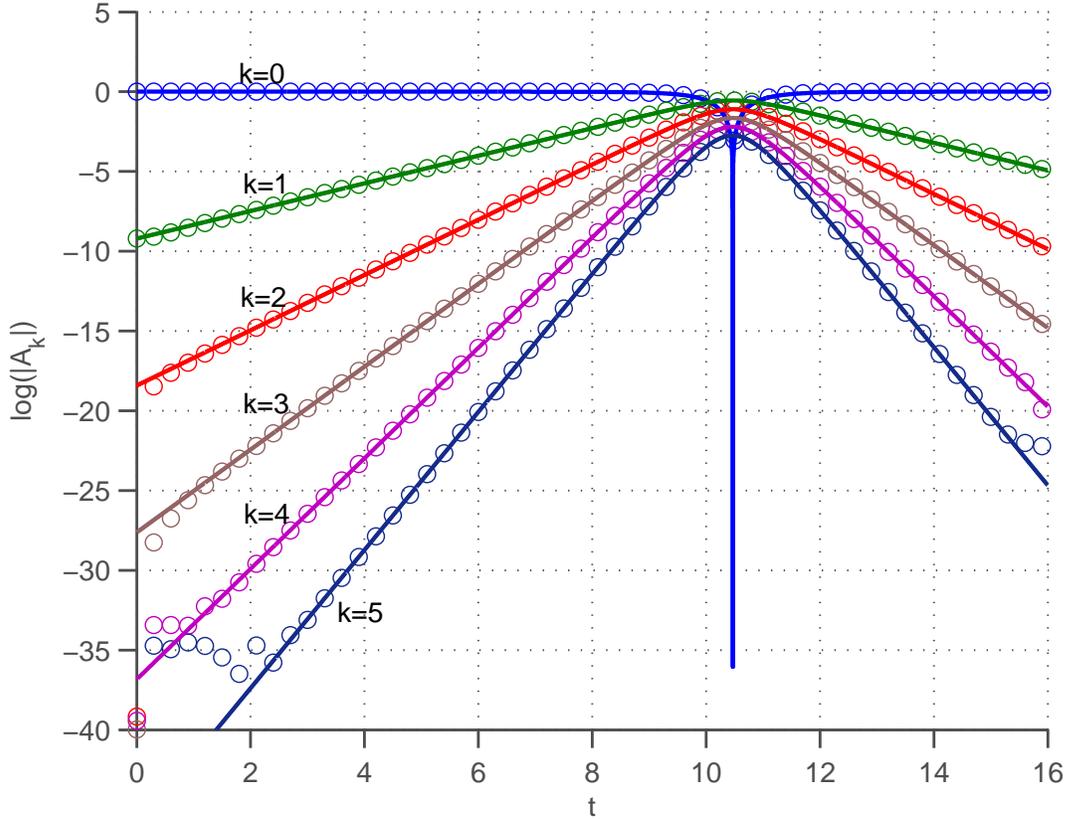}
\caption{ The growth of the $k=0-5$ Fourier amplitudes
as compared to the exact amplitudes (\ref{am}) and (\ref{amn}).
The numerical data are the same as those in Fig. \ref{kmode}.
The plunging vertical line depicts the total depletion of $A_0$ (=0) at $t=t_0$.
\label{emode}}
\end{figure}

 To see how exactly the Akhmediev breather forms from this
cascading scenario, we now compute the exact amplitudes
 $A_n(t)$ from the solution (\ref{ab}).
From Eq. (\ref{ab}), we have
\ba
 A_n(t)&=&\frac1{L}\int_0^L\psi(t,x)\cos(n\Omega x) dx,\nn\\
&=&\frac1{2\pi}\int_0^{2\pi}\biggl(1+ \frac{2(1-2a)\cosh(\ld t)+i\ld \sinh(\ld t)}{\sqrt{2a}\cos(y)-\cosh(\ld t)}\biggr)\cos(ny) dy,\nn\\
&=&\frac1{2\pi}\int_0^{2\pi} \biggl(1+\frac{2(1-2a)+i\ld \tanh(\ld t)}{\alpha\cos(y)-1} \biggr)\cos(ny)dy,
 \ea
where we have defined
$$\al=\sqrt{2a}/\cosh(\ld t)<1,$$
and where
\be
\frac1{2\pi}\int_0^{2\pi}\frac{\cos(ny)}{\alpha\cos(y)-1}dy=-\frac1{\sqrt{1-\al^2}}\biggl(\frac{1-\sqrt{1-\al^2}}{\al}\biggr)^n.
\la{intg}
\ee
Therefore, one has
\be
	 A_0(t)=1-\frac{2(1-2a)+i\ld \tanh(\ld t)}{\sqrt{1-\al^2}},
\la{am}
\ee
and for $n\ne 0$,
\be
 A_n(t)=-\frac{2(1-2a)+i\ld \tanh(\ld t)}{\sqrt{1-\al^2}}\biggl(\frac{1-\sqrt{1-\al^2}}{\al}\biggr)^{|n|}.
 \la{amn}
\ee
This derivation agrees with the original results of
Akhmediev and Korneev \cite{akh86} (up to an overall sign) for $a=1/4$, and with others \cite{ham11},
but not with the amplitudes given in Ref. \cite{akh11}.

Equation (\ref{amn}) means that all amplitudes are phase-locked to that of $A_1(t)$ and their
magnitudes simply decrease geometrically with increasing $n$:
\be
 A_n(t)=\biggl(\frac{1-\sqrt{1-\al^2}}{\al}\biggr)^{|n|-1}A_1(t).
 \la{amnn}
\ee
This affirms the cascading scenario, but the
Akhmediev breather goes further in asserting that all $|n|>1$ amplitudes
evolve in locked-step with $A_1(t)$ at {\it all times}. In 1981, Infeld \cite{inf81}
explained the success of his truncated three-wave model as due to the
``enslavement" of higher modes to the $n=1$ modes. This conjecture is
precisely confirmed by the Akhmediev breather and is the basis for
the qualitative success of all three-wave models \cite{inf81,tri91}.
This ``enslavement" is a remarkably simple mechanism of nonlinear evolution.

Note that
\be
	 A_0(0)=1-2\sqrt{1-2a}=1-\w
\la{amo}	
\ee
and for $n\ne 0$,
\be
 A_n(0)=-2\sqrt{1-2a}\biggl(\frac{1-\sqrt{1-2a}}{\sqrt{2a}}\biggr)^{|n|}.
 \la{amon}
\ee
Therefore, the depletion of the background is maximal, $A_0(0)=0$, only for $a=3/8$, $\w=1$.
We now plot $\ln(|A_{k}(t-t_0)|)$ in Fig. \ref{emode},
using the exact AB amplitudes of (\ref{am}) and (\ref{amn}).
The exact amplitudes match the numerical data perfectly.

When the time origin is shifted by $t\rightarrow t-t_0$, then for $t<<t_0$
$$\al\rightarrow 2\sqrt{2a}\e^{\ld (t-t_0)}\rightarrow 0$$ and
\be
 |A_n(t-t_0)|\rightarrow2\sqrt{1-2a}\biggl(\sqrt{2a}
 \e^{\ld (t-t_0)}\biggr)^{|n|}.
 \la{exsp}
\ee
Therefore, at any $a$ the Akhmediev breather will yield a growing triangular
spectrum, as predicted by the cascading instability, but with a known prefactor $C_n$.

\begin{figure}
\includegraphics[width=0.95\linewidth]{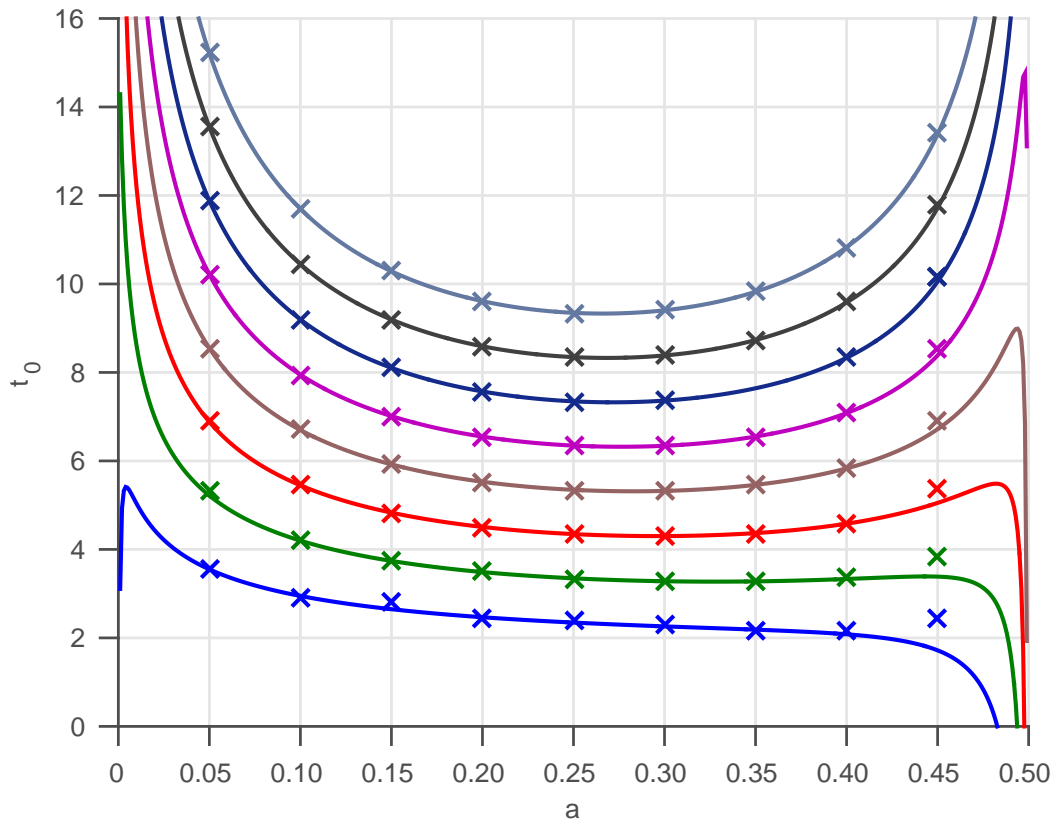}
\caption{ The corrected Akhmediev breather's first formation times as functions of $a$ for various
initial amplitudes parametrized as $A_1=\e^{-w}$. Crosses are numerical values; lines are analytical estimates
of (\ref{tzeroc}). From the top to the bottom are results corresponding to
$w=9-2$.
\label{formqa}}
\end{figure}

 \section{The formation time}

The excellent match in Fig. \ref{emode} between numerical data and theoretical results means that one
can track $\ln(|A_{1}(t-t_0)|)$ back to $t=0$, and 
set it equal to the initial amplitude,
\be
\ln A_1=\ln(|A_{1}(-t_0)|),
\ee
and directly determine $t_0$ from the initial amplitude! From Fig. \ref{emode},
for $A_1$ small it is clear that  $\ln(|A_{1}(t-t_0)|)$ is in the linearly growing region.
There is no need to use the full expression (\ref{amn}); the approximation (\ref{exsp})
\be
 |A_1(-t_0)|= \ld
 \e^{-\ld t_0}
 \la{exsp1}
\ee
is sufficient. One then has an analytical form for the formation time:
\be
t_0=-\frac{\ln(A_1/\ld)}{\ld}=-\frac{\ln(A_1)}{\ld}+\frac{\ln\ld}{\ld}.
\la{tzero}
\ee
Setting $A_1=10^{-4}$ gives
$
t_0=10.6352-0.1661=10.4691,
$
in exact agreement with the observed formation time of (\ref{tf}).

The analytical formation time (\ref{tzero}) is exact for $A_1\rightarrow 0$.
It is surprising to find that (\ref{tzero}) remains a good approximation at $a=3/8$ even
for $A_1$ as large as $\approx 0.1$. However, at larger values of $A_1$, there is not
enough time for the cascading process to build up to the triangular spectrum,
and the resulting evolution is no longer described by the Akhmediev breather.
Thus, in order for the Akhmediev breather and (\ref{tzero}) to be applicable,
the smaller the initial modulating amplitude, the better.

The above discussion for $t_0$ is only for the case of $a=3/8$, $\w=1$. It turns out that for $\w\ne 1$,
another correction is necessary. This is due to the fact that for $\w\ne 1$, the numerical
$\ln(|A_1(t)|)$ will start out either steeper or flatter than the slope $\ld$.
In general, starting with a finite $A_1\ne 0$, unless the initial $A_1$ is very small, 
$\ln(|A_1(t)|)$ is not described by 
the Akhmediev breather. But if we are using the straightline portion (\ref{exsp1}) of the Akhmediev breather to
determine $t_0$ approximately, then we must use a value of $|A_1|$ logarithmically higher or lower,
to match the slope. The correction, found empirically, takes a very simple form:

\begin{figure}
\includegraphics[width=0.95\linewidth]{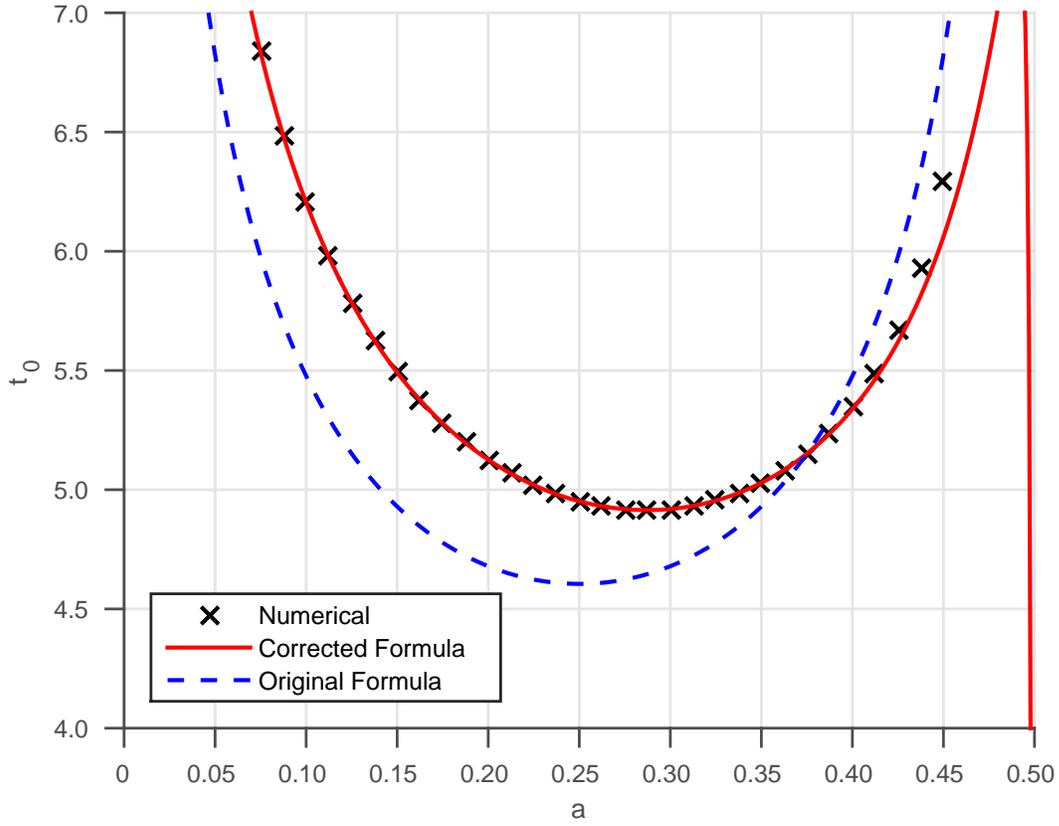}
\caption{Comparing the original formula (\ref{tzero}) and the corrected formula (\ref{tzeroc})
 for the formation time $t_0$ at $A_1$=0.01.
\label{apoh1}}
\end{figure}

\be
t_0=-\frac{\ln[A_1/(\ld\w)]}{\ld}=-\frac{\ln(A_1)}{\ld}+\frac{\ln\ld}{\ld}+\frac{\ln\w}{\ld}.
\la{tzeroc}
\ee
In Fig. \ref{formqa}, we compare the numerical formation time for various values of
the modulating amplitude $A_1$ and $a$.
At smaller values of $A_1$, $w=7,8,9$, the agreement is excellent even at $a>0.375$,
where there are more than one initial unstable modes. Naturally, the analytical estimates fail 
as one approaches $a=1/2$, where the AB scenario is not applicable anymore and the Peregrine soliton scenario takes over.

At $A_1=0.01$, we compare both (\ref{tzero}) and (\ref{tzeroc}) in Fig. \ref{apoh1} for all
values of $a$. The old formula (\ref{tzero}) is only correct at one point, $a=3/8=0.375$ or $\w=1$.
The corrected formula (\ref{tzeroc}) is in excellent agreement with numerical results,
even at this relatively large value of $A_1$.
If one interprets $t$ as $z$, the distance along the optical length, then $t_0$ corresponds to the
distance at which $A_0$ is maximally depleted. This is called the maximum compression distance
by Erkintalo {\it et al.} in Ref. \cite{erk11a}. They have also derived the formation time (\ref{tzero})
by an entirely different expansion method. Our use of the amplitude $|A_1(t)|$ from (\ref{exsp})
seemed more direct. Figure \ref{apoh1} is to be compared with Fig. 3 in Ref. \cite{erk11a}.

\begin{figure}
\includegraphics[width=0.95\linewidth]{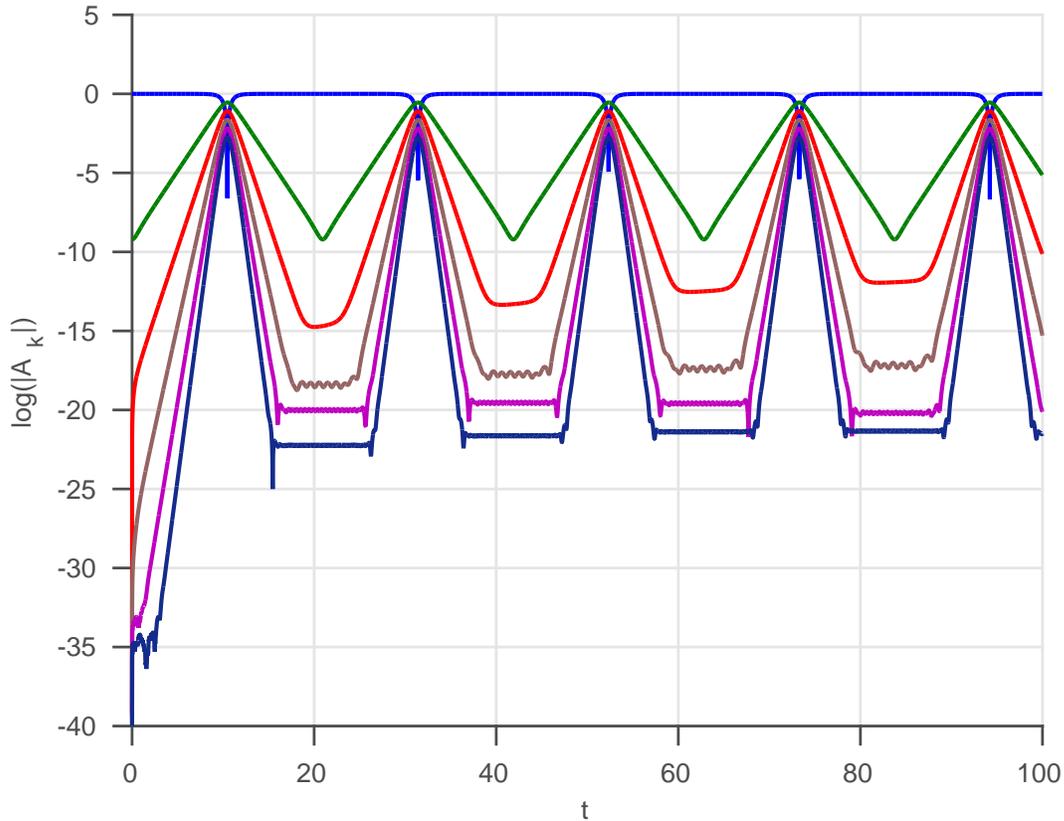}
\caption{ The evolution of the $k=0-5$ Fourier amplitudes (top to bottom)
beyond the first peak.
\label{allkt}}
\end{figure}

\section{ Fermi-Pasta-Ulam recurrence}

In Fig. \ref{allkt}, we show the continued evolution of the amplitudes
for the case of $a=3/8$ after the first peak.
One sees that $A_1$ decreases back to its initial value and repeats its initial growing
pattern. This is the celebrated Fermi-Pasta-Ulam \cite{fer65,ford92} recurrence of the nonlinear
Schr\"odinger equation, known from the early experimental work of Lake {\it et al.} \cite{lak77}
and the numerical calculations of Yuen and Ferguson \cite{yue78}. As discussed earlier,
in AB, all higher amplitudes should rise and fall in locked-step with $A_1$.
However, due to the algorithm's error and limited numerical precision, this locked-step is difficult
to maintain when the amplitudes are near zero.  ($A_2$ is specially difficult
here because it is neutrally stable with a zero growth rate. It tends to drift more than other modes.)
The exact AB is of no help in explaining this recurrence, since the exact wave function (\ref{ab})
only describes a single rise and fall of the breather. The key to understanding this recurrence is energy conservation.

\begin{figure}
\includegraphics[width=0.95\linewidth]{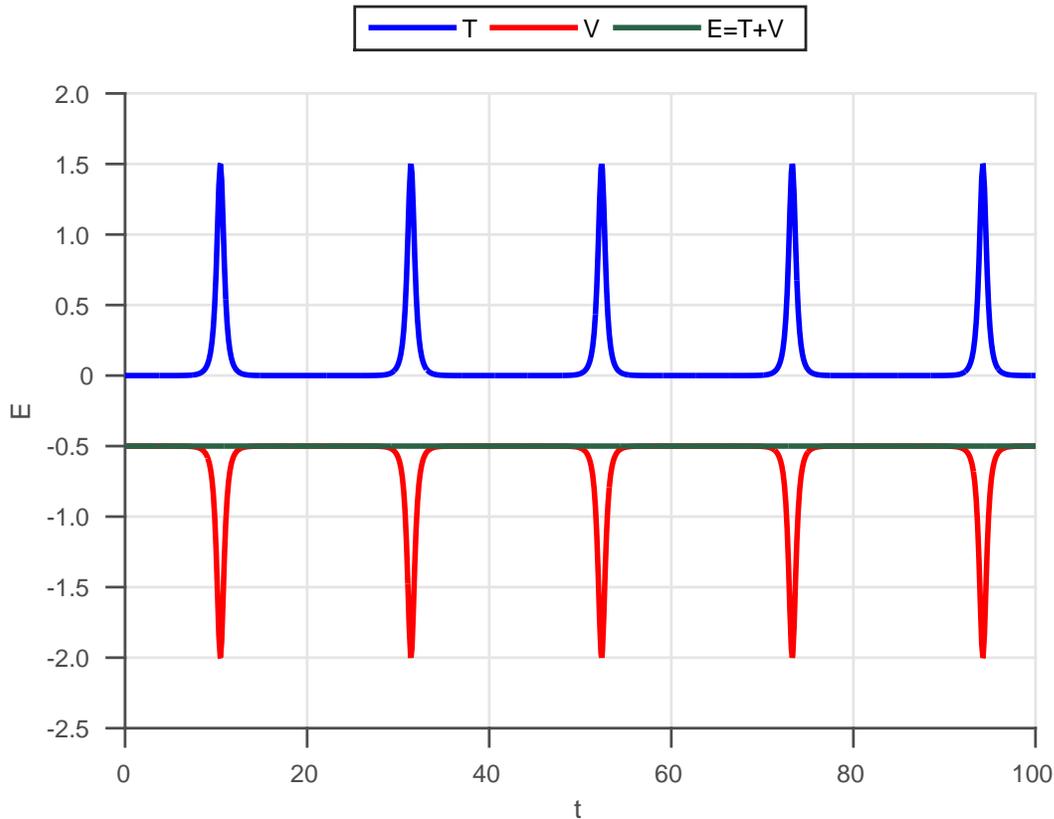}
\caption{ The kinetic energy (top) and the potential energy (bottom) of the nonlinear Schr\"odinger equation
with initial wave function (\ref{init}). The line nearly identitical to -1/2 is the total energy.
\label{eng}}
\end{figure}

In Fig. \ref{eng} we, plot the kinetic energy $T$, the potential energy $V$, and the total energy $E=T+V$
of the nonlinear Schr\"odinger equation (\ref{sch}) as a function of $t$, where $T$ and $V$ are
defined by
\ba
T&=&\frac1{L}\int_0^L\!\! dx\, \psi^*(t,x)(-\frac12\pa_x^2)\psi(t,x)=\sum_{k}\frac12 k^2|A_k(t)|^2,\nn\\
V&=&-\frac12\frac1{L}\int_0^L\!\! dx\, |\psi(t,x)|^4.
\nn
\ea
At the peak of the AB,
(\ref{amo}) and (\ref{amon}) give $A_0=0$  and
\be
|A_k|^2=\frac1{3^k},
\ee
and hence
\be
T=2\sum_{k=1}^{\infty}|A_k|^2\frac12 k^2=\sum_{k=1}^{\infty}\frac {k^2}{3^k}=\frac32.
\ee
This exlains why the kinetic energy peaks at 1.5.

One immediately recognizes that energy patterns in Fig. \ref{eng} are typical of a bound state collision with
a hard-wall potential, like that of a bouncing ball released from rest at a given height, 
then falling to the ground. When the ball hits the ground, its velocity and kinetic energy are at their
respective maxima. When the ball begins to bounce back elastically,
its velocity reverses direction and both its magnitude
and the kinetic energy decrease back to zero. When the ball reaches back its original height with zero kinetic
energy, it begins to fall again. Thus every kinetic energy peak is a moment of impact.
The FPU period is  therefore just the period of the bouncing ball, which is twice the time for it to fall to the ground.
Hence,
\be
t_{\rm FPU}=2t_0=2\biggl(-\frac{\ln[A_1/(\ld\w)]}{\ld}\biggr).
\la{fpu}
\ee
This is clearly seen in the first optical observation of the FPU recurrence \cite{sim01}.

\begin{figure}
\includegraphics[width=0.95\linewidth]{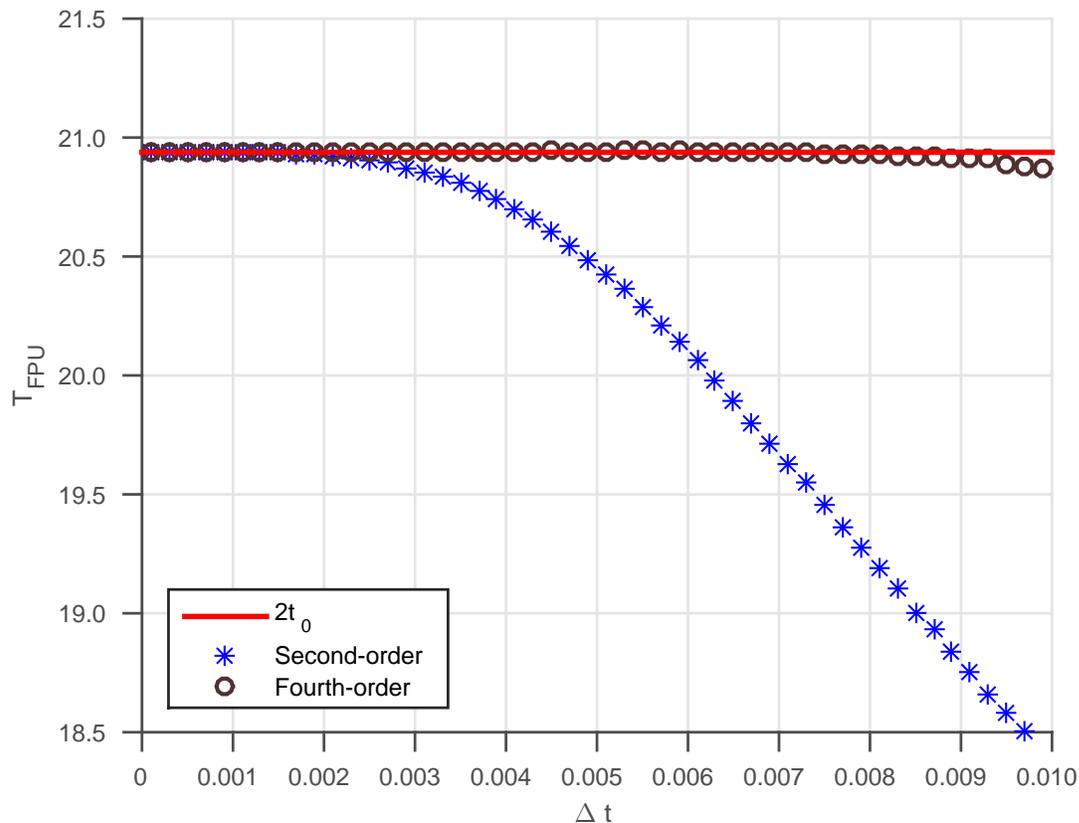}
\caption{ The Fermi-Pasta-Ulam recurrence period $t_{\rm FPU}$ as determined by a second (stars) and a fourth-order (circles) splitting scheme with $\Delta t=0.01$ to 0.0001 at $a=3/8$ for an initial amplitude of $A_1=10^{-4}$.
The horizontal red line is the predicted FPU period of $2t_0$.
\label{fput}}
\end{figure}

In the original observation of Yuen and Ferguson \cite{yue78}, recurrence in the nonlinear Schr\"odinger is liken to the work
of Fermi-Pasta-Ulam \cite{fer65}, because it was thought that energy is being distributed
from $A_0$ to infinite-many higher Fourier modes. If all these higher modes interact
{\it independently}, then the energy will  thermalize and impossible to reassemble
back to $A_0$. From AB, we now have a simple explanation of this recurrence:
{\it All} higher modes are locked-in, to rise and fall with $A_{\pm 1}(t)$ at all times.
There are therefore no infinite number of degrees of freedom to distribute energy, no entropy to destroy
time-reversal symmetry. Recurrence is just a matter of simple energy conservation for
basically two degrees of freedom, $A_0(t)$ and $A_1(t)$, similar to that of a two-body collision problem.

If one were to numerically determine the period of the bouncing ball accurately, then the ball
must be able to return to its original height accurately. In other words, energy conservation is paramount.
For the wave function (\ref{init}), the sum of kinetic and potential energy initially is
\be
E_0=A_1^2-\frac12 -4A_1^2+7A_1^4.
\ee
Since $A_1^4=10^{-16}$, the total energy is near the limit of double-precision. For $A_1<10^{-4}$, the
total energy would be beyond the limit of double-precision and cannot be accurately conserved numerically
(unless higher precision software is used).

To demonstrate the importance of energy conservation, we plot in Fig. \ref{fput}, the
FPU recurrence period using a second and a fourth-order splitting algorithm
with $\dt=0.01$ to 0.0001. With decreasing $\dt$, as the second algorithm is
increasingly more accurate, its numerical FPU period approaches the expected value of $2t_0$ from below,
and does not converge until $\dt<0.002$. However, the same converged value can
be obtained by using a fourth-order algorithm at step size as large as $\dt= 0.008$.
Also, as shown by the second order algorithm, when energy conservation is less accurate,
the FPU period tends to shorten.

\section{Higher-order modulation instability}

After understanding simple FPU recurrences in the last section, we can now tackle
the more complicated case of higher-order modulated instablility, with multiple initial unstable modes.

\begin{figure}
	\includegraphics[width=0.95\linewidth]{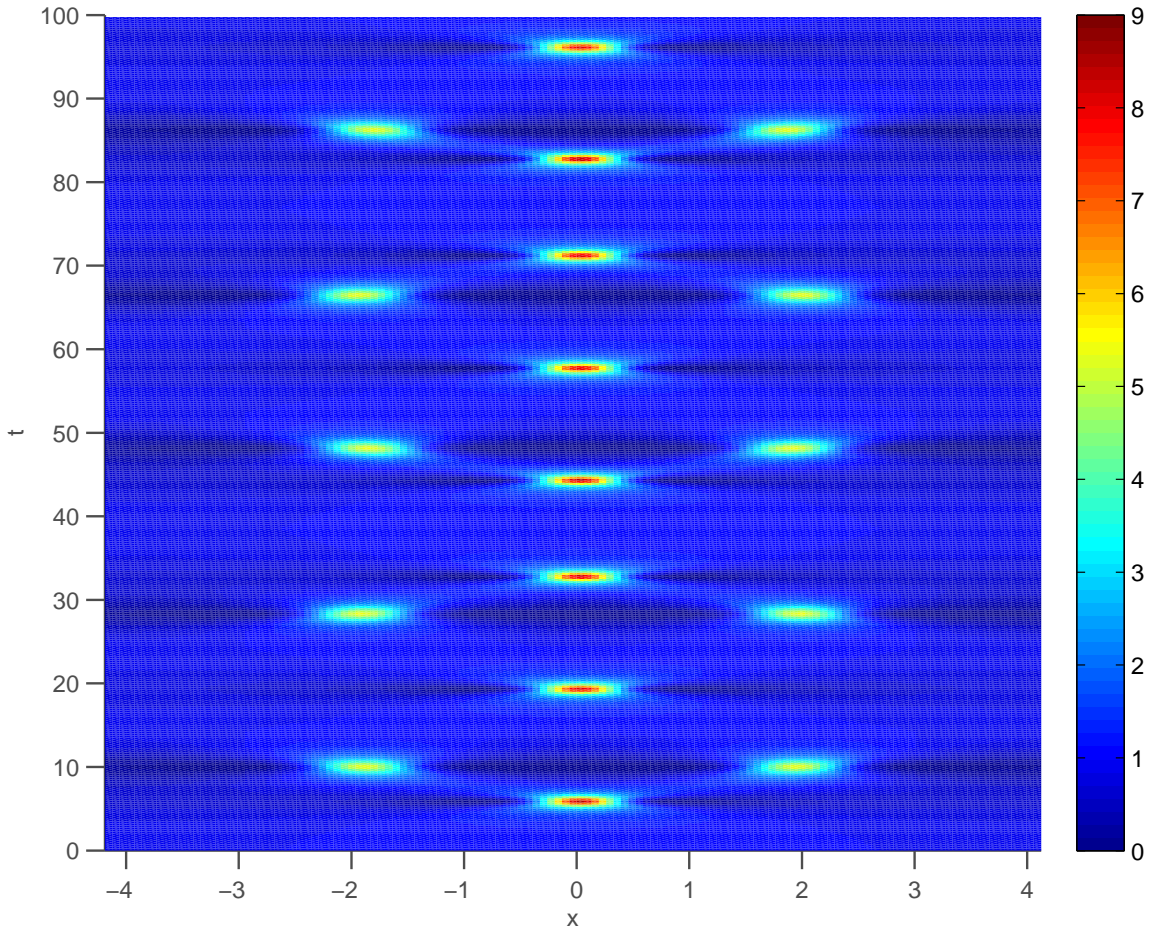}
	\caption{Density $|\psi(t,x)|^2$ plot for $a=55/128$.
		\label{q43}}
\end{figure}

\begin{figure}
	\includegraphics[width=0.95\linewidth]{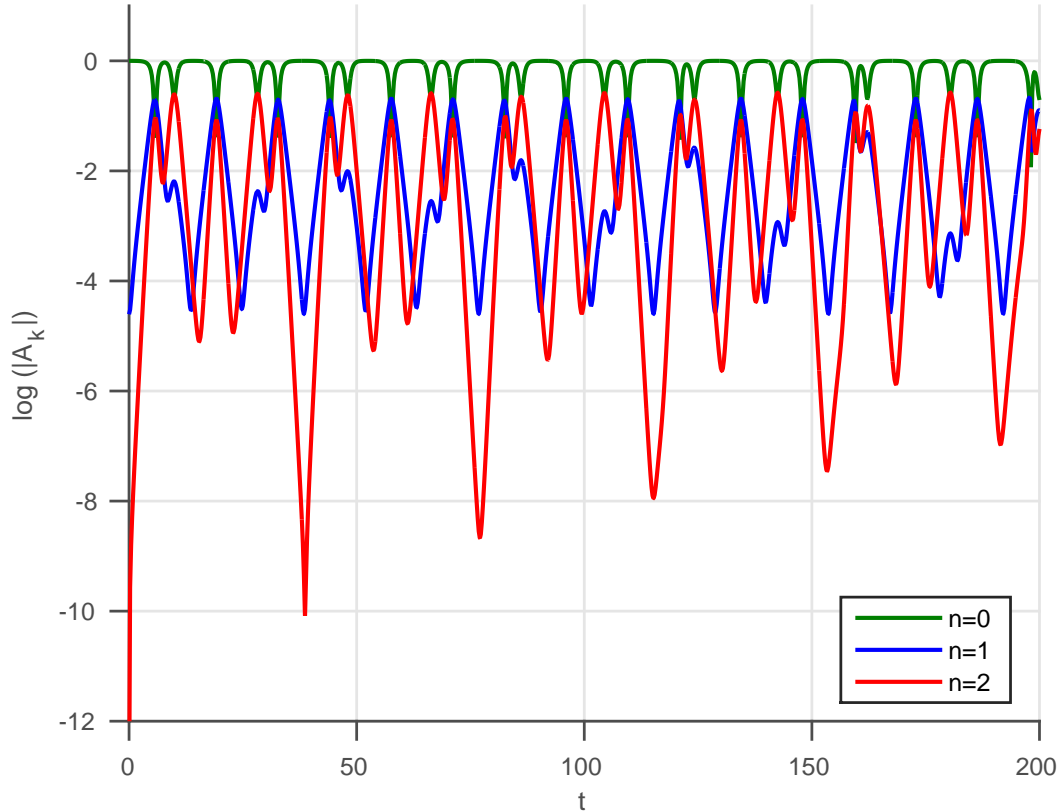}
	\caption{Evolving amplitude plot for $a=55/128$ with the
	super-recurrence period most clearly seen in the oscillation of the $n=2$ amplitude.
		\label{q43k}}
\end{figure}

Consider the case of $a=55/128=0.4296875$, where the fundamental wave number $\w_1=3/4$
and its first harmonic $\w_2=3/2$ are both unstable, according to the Bogoliubov spectrum,
with respective growth factors $\ld_1=0.6953$ and $\ld_2=0.9922$.
This is the case studied in Ref. \cite{erk11}. In Fig.\ref{q43}, we show the
resulting density plot when the initial modulating amplitude $A_1$ is $0.01$. A short simulation with $t<15$
would only show one breather near $t\sim5$ and two breathers near $t\sim10$. However, the long time simulation
of Fig. \ref{q43} reveals that there is ``super-recurrence" with the period $t_{FPU}\sim40$!
This is more clearly shown in Fig. \ref{q43k}, where for clarity, we only plotted the amplitudes of the 
first three modes. The longer period of $t_{FPU}\approx 38.4$ is clearly visible in the
oscillation of the $n=2$ amplitude.

Even for $A_1=0.01$, the cascading instability still enslaves all the higher modes,
including the unstable $n=2$
mode. This is because under the cascading instability, the $n=2$ mode will grow at a rate of
$2(0.6953)=1.3906$, which is faster than its own rate of  0.9922.
As a consequence, an Akhemdiev breather will form at
\ba
t_0&=&-\frac{\ln(A_1)}{\ld}+\frac{\ln\ld}{\ld}+\frac{\ln\w}{\ld}\nn\\
&=& 6.6233-0.5227-0.4138=5.6868.
\ea
This is in reasonable agreement with the observed value of $t_0=5.759$, given the
fact that $A_1=0.01$ is not small enough for the analytical formula to hold.
This time difference at $a\approx0.43$ is already noticeable in Fig. \ref{apoh1}.

\begin{figure}
	\includegraphics[width=0.95\linewidth]{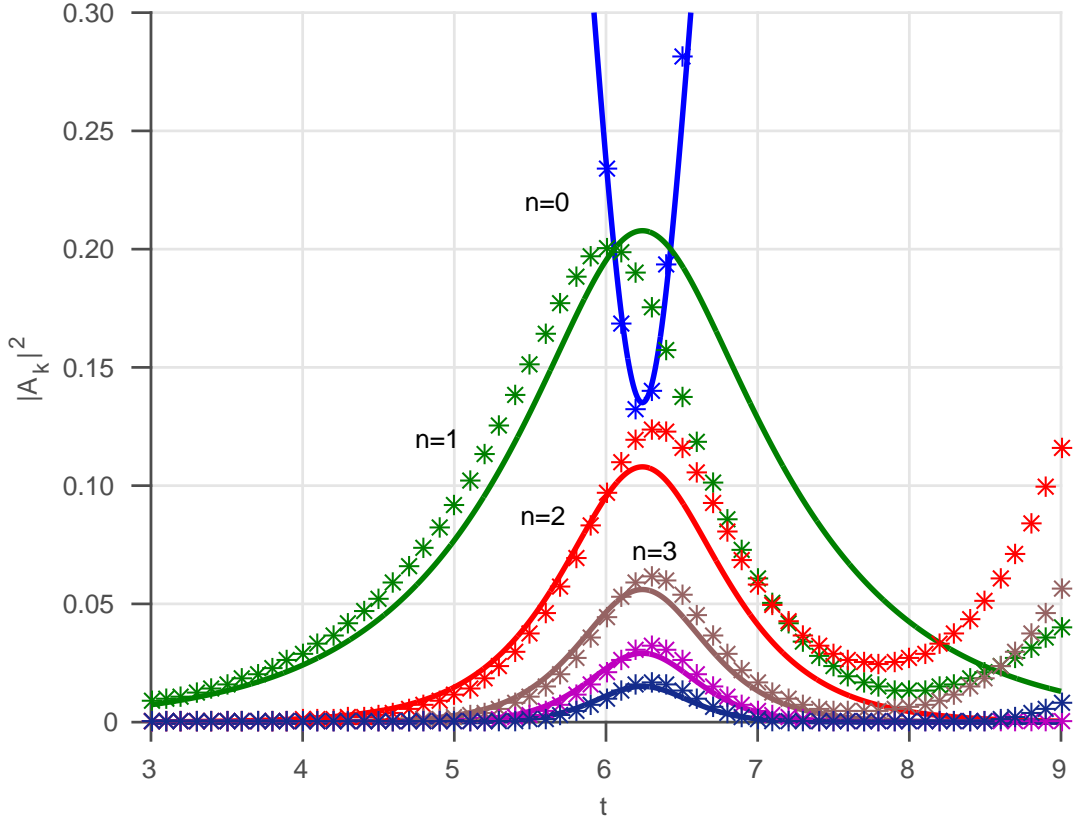}
	\caption{Comparing amplitude intensity of the Akhmediev breather (lines) 
   with numerical results (stars) for $a=0.45$. 
		\label{q45}}
\end{figure}

After the first peak, all amplitudes decline, as in an Akhemdiev breather, but since the $n=2$ mode is
intrinsically unstable, it starts to grow at its own rate while $A_1$ is still declining.
The result is the formation of a twin-peak breather near $t\approx 9.9$, where $n=2$ has the largest
amplitude. Wabnitz and Akhmediev \cite{wab10} have noted that this can be an efficient
way of transferring power from the pump or the background to the $n=2$ mode.
After this peak, the amplitdues decline and rise again, to form another Akhemdiev breather near $t\approx 19.2$. 
This peak is at mid-period, and the $A_1$ amplitude
retrace its step back to its starting value at the ``super-recurrence" period of $t_{FPU}\approx 38.4$.
This period is even more clearly seen in the $n=2$ amplitude, where it is the time at which
$A_2$ declines back to zero. 
It would be of interest to see whether this super-recurrence period can be seen
analytically in the solutions obtained by Darboux transformation in Ref. \cite{erk11}.

Because of its instability, the amplitude of the $n=2$ mode after the first peak is markedly different
from that of an Akhmediev breather. In Fig. \ref{q45}, we compare AB  at $a=0.45$ with
numerical results for $A_1=0.01$. This case has {\it three} unstable initial modes. 
The evolution before the first peak, because of the cascading instability, remain AB-like, 
however, after the peak,  numerical results for the intensity of the $n=1$ and $n=2$ mode
are below and above that of AB's profile respectively. This is to be compared with Fig. 3 of Ref. \cite{ham11}. 
Hammani {\it et al.} \cite{ham11} are correct in asserting that AB dynamics remained qualitatively
useful in describing various mode intensities in approaching the first peak. However, their Fig. 3
also clearly shows that their data after the intensity peak, while still in agreement with numerical
solutions of the nonlinear Schr\"odinger equation similar to our Fig. \ref{q45},
are no longer quantitatively described by the Akhmediev breather. Thus, while there are many areas where
AB can give an excellent account of light propagation and generation \cite{dul09},
AB dynamics is insufficient to describe higher-order modulation instabilities {\it beyond} the first peak.
With more than two unstable modes, that dynamics fast becomes chaotic.
\section{Conclusions}

In this work, we have exposed in detail the anatomy of producing the Akhmediev breather,  that it
is basically the result of a cascading instability, which enslaved all higher modes to evolve
in locked-step with the $n=1$ mode. This at once makes plain that FPU recurrence is just
a necessary consquence of energy conservation. By giving an analytical formula for the breather's first 
formation time beyond that of AB, we have also derived an accurate
analytical estimate of the FPU period. In cases of higher-order modulation instability, where
there are multiple unstable initial modes, we showed that due to the interplay between and
among various unstable modes, super-recurrences are possible. However, such recurrences are beyond
the simple description of AB (\ref{ab}) and may require the use of Darboux
transformations \cite{erk11} to give an analytical account of such periodicities. 

Because of the cascading instability, we were led to plot {\it not} the mode intensities $|A_n|^2$
themselves, but $\ln(|A_n|)$. As evident from Figs. \ref{kmode}, \ref{emode}, \ref{allkt} and \ref{q43k},
these log-plots of amplitudes give the clearest description of the nonlinear evolution of ABs of
the Schr\"odinger equation. This will be equally useful for understanding other
modulation-instability related nonlinear evolutions. Further work is necessary to understand the
higher-order MI of ABs, as well as the emerging complex dynamics associated with MI of Peregrine 
and of KMBs.

\begin{acknowledgments}
This publication was made possible by NPRP GRANT \#5-674-1-114 
from the Qatar National Research Fund (a member of Qatar Foundation).
MRB acknowledges support by the Al-Sraiya Holding Group. 
\end{acknowledgments}

\end{document}